\documentclass[preprint,12pt]{elsarticle}
\usepackage{epsf,epsfig}
\usepackage[psamsfonts]{amssymb}
\usepackage{amsmath}
\usepackage{bm}
\usepackage{graphicx}
\usepackage{graphicx,epstopdf}
\usepackage{caption}
\usepackage{subcaption}
\newcommand{\bfsfG}{\mbox{\sffamily\bfseries{G}}}
\begin{document}
\begin{frontmatter}
\title{ Spontaneous emission of a moving atom in the presence of magnetodielectric material: A relativistic approach}
\author{Fariba Shafieiyan$^1$, Ehsan Amooghorban\footnote{Ehsan.amooghorban@sci.sku.ac.ir}$^{1,2,3}$, and Ali Mahdifar$^{1,2,4}$}
\address{$^1$  Department of Physics, Faculty of Basic Sciences, Shahrekord University, P.O. Box 115, Shahrekord 88186-34141, Iran.}
\address{$^2$  Photonic Research Group, Shahrekord University, Shahrekord 88186-34141, Iran.}
\address{$^3$  Nanotechnology Research Center, Shahrekord University, Shahrekord 88186-34141, Iran.}
\address{$^4$  Department of Physics, Faculty of Science, University of Isfahan, Hezar Jerib, Isfahan 81746-73441, Iran.}
\begin{abstract}
In this paper, based on a canonical quantization scheme, we study the effect of the relativistic motion of an excited atom on its decay rate in the presence of absorbing and dispersive media. For this purpose, we introduce an appropriate Lagrangian and describe the center-of-mass dynamical variables by the Dirac field. We obtain the Hamiltonian of the system in a multipolar form and calculate the motion equations of the system in the Schr\"{o}dinger picture.
We find that the decay rate and the quantum electrodynamics level shift of the moving atom can be expressed in terms of the imaginary part of the classical Green tensor and the center-of-mass velocity of the atom.
\end{abstract}
\end{frontmatter}
\section{Introduction}
One of the most fundamental phenomena in quantum optics is the spontaneous emission caused by the inevitable
interaction of an excited atom with the vacuum-quantized electromagnetic field and/or the
reaction of the atom to its own radiation field~\cite{Milonni1994}. This process was first formulated theoretically by Dirac in 1927 and further by Weiskopff in 1930~\cite{Milonni1994}.
In order to study the spontaneous emission, usually the atom is assumed to be in rest, which leads to difficulties due to Heisenberg's
uncertainty relation~\cite{Guo2008}. In recent years there has been an increasing number of papers on the role of the center-of-mass motion in the process of spontaneous emission~\cite{Wilkens1994,Boussiakou2002,Cresser2003}, Abraham-Minkowski-controversy~\cite{Hinds2009,Barnett2010a,Barnett2010b,Milonni2010},
Aharonov-Bohm-type phase shifts~\cite{Leonhardt1998,Leonhardt1999,Horsley2005} and many important effects and applications associated with atomic motion in atom optics, laser cooling, trapping, and isotope separation experiments~\cite{Ashkin2006}.

By taking into account the so-called R\"{o}ntgen term in the atom-field interaction, Wilkens evaluated the velocity dependence of
the spontaneous decay rate of an atom which moves in free
space with a constant velocity $v$ to lowest order of $v/c$~\cite{Wilkens1994}.
Later, Boussiakou \textit{et al} used a rigorous canonical formalism in which the center-of-mass dynamics of the
atom is explicitly included and calculated the spontaneous decay of a moving excited atom in free space~\cite{Boussiakou2002}.
They showed that, irrespective of the orientation of the atomic dipole with respect to the
direction of motion, the decay rate of the atom from the point of view of an observer in the laboratory frame is in agreement with special relativity. This result has been confirmed by an alternative but less general approach based on the basic principles of special relativity, physical processes associated with amoving electric dipole and the Doppler shift~\cite{Cresser2003}. As a matter of fact, for more realistic cases atoms are not in free space, but move near the material media. Therefore, this is not a practical assumption for most real world applications.
%As a matter of fact, in practice the atom is not in free space and moves in material media.

It is well known that the presence of material media can change the structure of the fluctuating field of the
vacuum. Consequently, the spontaneous emission rate can be modified if
the atom moves with uniform nonrelativistic
speed near materials of different composition and shape.
In Refs.~\cite{Scheel2009,Barton2010,Lannebère2017}, authors considered a more realistic case and studied the spontaneous emission and the friction force
experienced by an atom moving with uniform nonrelativistic velocity parallel to a dielectric surface.
%
%the friction force and spontaneous emission by a two-level atom interacting with a dispersive dielectric in relative motion were studied.
%
However, the question that naturally arises in this context is on the emission process
occurring when an atom moves in absorbing magnetodielectric material relativistically.
It is expected that the relativistic motion of the atom affects the
radiative properties of the atom. The present paper is intended to respond this question.
Our work extends previous works on the spontaneous decay of the moving excited atom in free space~\cite{Boussiakou2002,Cresser2003},
to the relativistic motion in the presence of the material media.
%Here we would like to consider a more realistic case and
%

As a first step in studying the relativistic dynamics of a moving atom in the presence of absorbing media one has to provide the quantization of the electromagnetic field.
%The first step in studying the relativistic dynamics of a moving atom in the presence of absorbing media is the problem of quantization of the electromagnetic field.
Generally speaking, there are two approaches to quantize the electromagnetic field in the presence of material media: phenomenological and canonical approach.
In this paper, we follow the canonical methods as presented in~\cite{Amoghorban2010,Amooghorban2011,Morshed2016}. More details concerning this rigorous canonical
approach can be found in references~\cite{Amoghorban2010,Amooghorban2011,Morshed2016,Huttner1992,Jeffers1996,Suttorp2004,Amooshahi2009,Philbin2010}.

The paper is organized as follows.
%the satisfactory theory for exact description of spontaneous emission requires quantization of the radiation field,
In Sec.~\ref{Basic equations for non-relativistic dynamics}, we present a canonical quantization of the electromagnetic field interacting with moving charge particles in the presence of an isotropic, inhomogeneous and absorbing magnetodielectric medium.
%We perform our calculations within a Lagrangian formalism
%
We start from a convenient Lagrangian and obtain the canonical momenta and the Hamiltonian of the combined system.
We apply this approach to the case of two non-relativistic particles of opposite charges which form an atomic system.
On the Hamiltonian, we perform a unitary transformation and derive nonrelativistic multipolar Hamiltonian in
the electric-dipole approximation.
In Sec.~\ref{Sec:Relativistic dynamics}, we generalize the formalism by introducing the Dirac field to describe the relativistic motion of the atom.
%
%The evolution of the atomic system is determined by the
%Schrödinger equation, .
%Then, by considering the atom's external and internal degrees of freedom the evolution of the system is determined by the Schrödinger equation.
%
In Sec.~\ref{Sec:Radiative properties of a moving atom}, we examine the time evolution of the atomic system by treating the atom's external and internal
degrees of freedom on the same quantum footing in the Schr\"{o}dinger picture.
In this section, the decay rate and the quantum electrodynamics level shift of the two-level atom, that moving with relativistic velocity near dissipative media, are explicitly evaluated.
%It is found that the decay rate and the quantum electrodynamics level shift of the atom, that moving with relativistic velocity near dissipative media, are expressed in terms of the imaginary part of the classical Green tensor and the center-of-mass velocity of the atom before and after the emission of photon.
%
Finally, the main results are summarized in Sec.~\ref{Sec:conclusions}.

\section{Basic equations for non-relativistic dynamics}\label{Basic equations for non-relativistic dynamics}
Let us consider a system composed of charged particles, the electromagnetic field, an absorbing and dispersive magnetodielectric medium and the interactions between them. Since, for short particle-medium separations the macroscopic description of the medium is not justified, we assume that the charged particles are placed in the free space and well separated from the medium. The Lagrangian of the whole system is written as follows~\cite{Amoghorban2010,Amooghorban2011,Morshed2016,Huttner1992,Suttorp2004,Amooshahi2009,Philbin2010}
\begin{equation} \label{Lagrangian}
L = L_q + L_{em} + L_m + L_{int},
\end{equation}
where
\begin{equation} \label{Lagrangian of particles}
L_q = \frac{1}{2}\sum_\alpha m_\alpha{\dot{\bf r}}^2_\alpha (t),
\end{equation}
is the Lagrangian for the charged particles with masses ${m_\alpha }$, charges ${e_\alpha }$ and the position vector ${{\bf{r}}_\alpha}$, and the Lagrangian of the electromagnetic field, ${L_{em}}$, is given by ${L_{em}} = \frac{1}{2}\int d^3{\bf r} \big({\varepsilon _0}\,{{ { \pmb{\mathcal E}}}^2}({\bf r},t) - \frac{\pmb{\cal B}^2 ({\bf r},t)}{\mu _0}\big)$. Here, the electric and magnetic fields can be defined in terms of the vector potential ${\bf{A}}$ and the scalar potential $\varphi$ as $\pmb{\mathcal E} = - \nabla \varphi - \frac{\partial {\bf A}}{\partial t}$ and $\pmb{\mathcal B} = \nabla \times {\bf A}$, respectively. In the Coulomb gauge, respectively, $- \nabla \varphi $ and $- \frac{\partial {\bf A}}{\partial t}$ are  related to the longitudinal part $\pmb{\mathcal E}^\|$ and the transverse part $\pmb{\mathcal E}^\bot$ of the total electric field $\pmb{\mathcal E}$.

The third term in Eq.~(\ref{Lagrangian}), ${L_m}$, denotes the magnetodielectric medium part as
\begin{equation}\label{Lagrangian of magnetodielectric material}
L_m = \frac{1}{2} \int d^3{\bf r}\int_0^\infty d\omega \,\Big[{\dot{\bf X}}_\omega ^2 ({\bf r},t)+{\dot{\bf Y}}_\omega ^2 ({\bf r},t) - {\omega ^2} \big({\bf X}_\omega ^2 ({\bf r},t)+{\bf Y}_\omega ^2({\bf r},t) \big) \Big].
\end{equation}
Here, the medium is modeled by two independent sets of harmonic oscillators characterized by means of two medium fields ${{\bf{X}}_\omega }$ and ${{\bf{Y}}_\omega }$. This scheme is based on Hopfield's microscopic model~\cite{Hopfield1996}, which provide the dissipation of the energy as well as the polarizability and the magnetizability characters of the medium.

Finally, the interaction part of the Lagrangian~(\ref{Lagrangian}) is given by
\begin{eqnarray}\label{Interaction lagrangian}
L_{int} &=& \sum_\alpha \big[{e_\alpha {\dot {\bf r}}_\alpha \cdot {\bf A}({\bf r}_\alpha ,t) - e_\alpha \varphi ({\bf r}_\alpha ,t)}\big]\\
&&+\int d^3{\bf r}\,({\bf P}({\bf r},t) \cdot \pmb{\cal E}({\bf r},t)+{\bf M}({\bf r},t) \cdot \pmb{\cal B}({\bf r},t)),\nonumber
\end{eqnarray}
where the terms in the first line describe the interaction of charged particles with the electromagnetic field, and those in the second line represent the interaction between the electromagnetic field and the material fields with the polarization vector ${\bf{P}}$ and the magnetization vector ${\bf{M}}$, respectively. Polarization and magnetization vectors can be, respectively, expressed in terms of the electric coupling function, ${g_e}$, and the magnetic coupling function, ${g_m}$, as follows
\begin{equation}\label{polarization vector}
{\bf P}({\bf r},t) = \int_0^\infty d\omega g_e({\bf r},\omega){\bf X}_\omega({\bf r},t),
\end{equation}
\begin{equation}\label{magnetization vector}
{\bf M}({\bf r},t) = \int_0^\infty d\omega g_m({\bf r},\omega ){\bf Y}_\omega({\bf r},t).
\end{equation}
We will see later that the dielectric permeability and the magnetic permittivity of the medium can be naturally
expressed in terms of these coupling functions. To simplify the calculations, without loss of generality, we assume that the medium is isotropic. Therefore, the coupling functions ${g_e}$ and ${g_m}$ are both scalars, but take on tensor forms when the medium is anisotropic~\cite{Morshed2016}.

From the Lagrangian density~(\ref{Lagrangian}), the canonical conjugate momenta associated to each dynamical
variables can be obtained as
\begin{equation}\label{conjugate momentum of particle}
{\bf p}_\alpha(t) = \frac{\partial L}{\partial {\dot {\bf r}}_\alpha} = m_\alpha {\dot {\bf r}}_\alpha + e_\alpha {\bf A}({\bf r}_\alpha ,t),
\end{equation}
\begin{equation}\label{conjugate momentum of field}
- \varepsilon_0\pmb{\cal E}^ \bot ({\bf r},t) = \frac{\delta L}{\delta{\dot {\bf A}}({\bf r},t)} =\varepsilon_0\dot{{\bf A }} ({\bf r},t),
\end{equation}
\begin{equation}\label{conjugate momentum of X}
{\bf Q}_\omega ({\bf r},t) = \frac{\delta L}{\delta {\dot {\bf X}}_\omega ({\bf r},t)} = {\dot {\bf X}}_\omega({\bf r},t) +g_e({\bf r},\omega ){\bf A}({\bf r},t),
\end{equation}
\begin{equation}\label{conjugate momentum of Y}
{\bf \Pi}_\omega ({\bf r},t) = \frac{\delta L}{\delta {\dot {\bf Y}}_\omega ({\bf r},t)} = {\dot {\bf Y}}_\omega ({\bf r},t).
\end{equation}
To describe the system quantum mechanically, we follow the standard canonical quantization procedure and impose between the variables and their canonical conjugates, which are now operators on the Hilbert space, the following commutation relations
\begin{equation}\label{commutation relation of particle}
[{{\hat {\bf r}}_\alpha }(t)\,\,\,\,{{\hat {\bf p}}_\beta }(t)] = i\hbar{\delta _{\alpha \beta }},
\end{equation}
\begin{equation}\label{commutation relation of field}
[{\hat {\bf A}}({\bf r},t)\,\,\,\, -{\varepsilon _0}\,{\hat{ \pmb{\cal E}}^\bot }({\bf r'},t)] = i\hbar{\delta ^\bot }({\bf r} - {\bf r'})\,,
\end{equation}
\begin{equation}\label{commutation relation of X}
[{{\hat {\bf X}}_\omega }({\bf r},t)\,\,\,\,{{\hat {\bf Q}}_{\omega '}}({\bf r'},t)] = i\hbar \delta ({\bf r} - {\bf r'})\delta (\omega - \omega '),
\end{equation}
\begin{equation}\label{commutation relation of Y}
[{{\hat {\bf Y}}_\omega }({\bf r},t)\,\,\,\,{{\hat {\bf{\Pi }}}_{\omega '}}({\bf r'},t)] = i\hbar \delta ({\bf r} - {\bf r'})\delta (\omega - \omega '),
\end{equation}
whereas all the other commutators of the canonical variables vanish.
% In the presence of charged particles, the longitudinal component of the electric field is written as ${\bf E}^\|=-\frac{{\bf P}^\|}{\varepsilon_0}$.
With the help of the above canonical momenta, we can now derive the hamiltonian of the system. After some algebra it reads
\begin{eqnarray}\label{Hamiltonian}
\hat H &=& \sum\limits_\alpha \frac{[{{\hat {\bf p}}_\alpha }-e_\alpha {\hat {\bf A}} ({\hat {\bf r}}_\alpha , t)]^2}{2 m_\alpha} + \frac{1}{2}\int {{d^3}{\bf r}} [{\varepsilon _0}\,{\hat {\pmb{\cal E}}^{\bot 2}}({\bf r},t) + \frac{{{\hat{\pmb{\cal B}}^2}({\bf r},t)}}{{{\mu _0}}}]\nonumber \\
&+& \frac{1}{2}\int {{d^3}{\bf r}} \int_0^\infty d\omega \,[{\hat {\bf Q}}_\omega ^2({\bf r},t) + {\omega ^2}{\hat {\bf X}}_\omega ^2({\bf r},t)]\nonumber \\
&+& \frac{1}{2}\int {{d^3}{\bf r}} \int_0^\infty d\omega \,[{\hat {\bf{\Pi }}}_\omega ^2({\bf r},t) + {\omega ^2}{\hat {\bf Y}}_\omega ^2({\bf r},t)] \\
&-& \int d^3{\bf r} [{\hat {\bf M}}({\bf r},t) \cdot {\hat{\pmb{\cal B}}}({\bf r},t) + {\hat {\dot {\bf{P}}}}({\bf r},t) \cdot {\hat {\bf A}}({\bf r},t)]\nonumber \\
&-& \frac{1}{2}\int d^3{\bf r} \int_0^\infty d\omega\,{g_e}({\bf r},\omega ){\hat {\bf A}}^2({\bf r},t)+w_{coul}, \nonumber
\end{eqnarray}
where the Coulomb energy, ${w_{coul}}$, is due to the interactions between the charged particles, the charged particles and the polarization charges, and the interactions between the polarization charges, and is defined as follows~\cite{Morshed2016}
\begin{eqnarray}\label{{w_{coul}}}
{w_{coul}} &=& \frac{1}{2}\int d^3{\bf r} \,{\hat \rho} _{A}({\bf r}){{\hat \varphi} _A}({\bf r}) + \int d^3{\bf r} \,{\hat \rho} _{A}({\bf r}) {\hat \varphi}_P ({\bf r})\nonumber \\
&+& \frac{1}{2}\int d^3{\bf r} \,{\hat \rho}_P({\bf r}){\hat \varphi}_P ({\bf r}).
\end{eqnarray}
Here, ${\hat \rho} _A ({\bf{r}}) = \sum_\alpha{{e_\alpha }\delta ({\bf{r}} - {{\hat {\bf r}}_\alpha })}$, ${\hat \rho} _P ({\bf{r}}) = -  \nabla \cdot {\hat {\bf{P}}}({\bf{r}})$,  are, respectively, the charge density, the polarization charge density, and the scalar potential which are attributed to the external and the polarization charges, according to: ${{\hat {\varphi}} _{A(P)}} = \frac{1}{{4\pi {\varepsilon _0}}}\int d^3{\bf{r'}} \frac{{{{\hat {\rho}}_{A(P)}}({\bf{r'}})}}{{|{\bf{r}} - {\bf{r'}}|}}$. Due to the presence of the external charged particles, the longitudinal and transverse components of the electric field can be written respectively as ${\hat{\pmb{\cal E}}}^\|={\hat {\bf E}}^\|-\nabla {\hat{ \phi}}_A$ and ${\hat{\pmb{\cal E}}}^\bot={\hat{\bf E}}^\bot$, wherein ${\hat {\bf E}}^\|=-\nabla{\hat {\varphi}}_P=-{\hat {\bf P}}^\|/\varepsilon_0$~\cite{Morshed2016,Dung2003}. Here, the total electric filed in the absence of the charged particles is denoted by ${\hat {\bf E}}$. Accordingly, the induction field ${\hat{ \bf B}}$ in the presence and absence of the
charged particles is given by ${\hat{\pmb{\cal B}}}={\hat{\bf B}}$~\cite{Morshed2016}.

In the Heisenberg picture, the time-evolution of operators are given by the Heisenberg equation.
By using the Hamiltonian~(\ref{Hamiltonian}), the equations of motion for the material fields ${\hat {\bf{X}}}_\omega $ and ${\hat {\bf{Y}}}_\omega $ are obtained as
\begin{equation}\label{motion equations of X}
{\ddot{\hat{\bf X}}}_\omega ({\bf r},t) + {\omega ^2}{{\hat {\bf X}}_\omega }({\bf r},t) = {g_e}({\bf r},\omega ) \cdot {\hat{\pmb{\cal E}}}({\bf r},t),
\end{equation}
\begin{equation}\label{motion equations of Y}
{\ddot{\hat {\bf Y}}}_\omega ({\bf r},t) + {\omega ^2}{{\hat {\bf Y}}_\omega }({\bf r},t) = {g_m}({\bf r},\omega ) \cdot {\hat{\bf B}}({\bf r},t).
\end{equation}
By solving these equations and substituting their solutions into Eqs.~(\ref{polarization vector}) and~(\ref{magnetization vector}),
%the dielectric polarization and the magnetic magnetization vectors of the matter are rewritten as follows
we derive
\begin{equation}\label{new polarization vector}
{\hat {\bf P}}({\bf r},t) = {\varepsilon _0}\int_0^\infty dt'\,{\chi _e}({\bf r},t - t'){\hat{\pmb{\cal E}}}({\bf r},t)\, + {{\hat {\bf P}}^N}({\bf r},t),
\end{equation}
\begin{equation}\label{new magnetization vector}
{\hat {\bf M}}({\bf r},t) = \mu _0^{ - 1}\int_0^\infty dt'\,{\chi _m}({\bf r},t - t'){\hat {\bf B}}({\bf r},t) + {{\hat {\bf M}}^N}({\bf r},t).
\end{equation}
where ${\chi _e}$ and ${\chi _m}$ are the electric and magnetic susceptibilities of the medium. In terms of the electric and magnetic coupling functions,  ${g_e}$ and ${g_m}$, these expressions take the form
\begin{equation}\label{electric susceptibility}
\chi _e({\bf r},t) = \theta (t) \frac{1}{{\varepsilon _0}}\int_0^\infty d\omega \,g_e^2({\bf r},\omega )\frac{{\sin \omega t}}{\omega },
\end{equation}
\begin{equation}\label{magnetic susceptibility}
\chi _m({\bf r},t) = \theta (t) {\mu _0} \int_0^\infty d\omega \,g_m^2({\bf r},\omega )\frac{{\sin \omega t}}{\omega },
\end{equation}
where $\theta (t)$ is the Heaviside function.

In Eqs.~(\ref{new polarization vector}) and (\ref{new magnetization vector}), ${{\hat {\bf{P}}}^N}$ and ${{\hat {\bf{M}}}^N}$ are, respectively, the noise polarization and the noise magnetization operators. It is worth nothing that these noise operators form a Langevin noise current  ${{\hat{\bf J}}^N}({\bf{r}},\omega ) = - i\omega {{\hat{\bf P}}^N}({\bf{r}},\omega ) + \nabla  \times {{\hat{\bf M}}^N}({\bf{r}},\omega )$~\cite{Buhmann2007}. These operators can be expressed in terms of the homogeneous part of the solution of Eqs.~(\ref{motion equations of X}) and (\ref{motion equations of Y}) as follows:
\begin{equation}\label{noise polarization}
{\hat {\bf P}}^N({\bf r},t) = \int_0^\infty d\omega \,{g_e}({\bf r},\omega )\,{\hat {\bf X}}_\omega ^N({\bf r},0),
\end{equation}
\begin{equation}\label{noise magnetization}
{\hat {\bf M}}^N({\bf r},t) = \int_0^\infty d\omega \,{g_m}({\bf r},\omega )\,{\hat {\bf Y}}_\omega ^N({\bf r},0),
\end{equation}
where ${\hat {\bf{X}}}_\omega ^N({\bf{r}},0) = {{\dot{\hat {\bf X}}}_\omega }({\bf{r}},0)\frac{{\sin \omega t}}{\omega } + {{\hat {\bf{X}}}_\omega }({\bf{r}},0)\cos\omega t$ and ${\hat {\bf{Y}}}_\omega ^N({\bf{r}},0) = {{\dot{\hat {\bf Y}}}_\omega }({\bf{r}},0)\frac{{\sin \omega t}}{\omega } + {{\hat {\bf{Y}}}_\omega }({\bf{r}},0)\cos\omega t$.
The equation of motion for the variables of the electromagnetic field follows by deriving the Heisenberg equation for the electric field ${\hat {\bf E}}$.
Using the solution of Eqs.~(\ref{motion equations of X}) and~(\ref{motion equations of Y}) for the medium fields, one can easily show that the wave
equation governing the electric field is obtained as
%By substituting the solution of Eqs.~(\ref{motion equations of X}) and~(\ref{motion equations of Y}) for the medium fields into this equation, a second-order differential equation for ${\bf E}$ is obtained
%
\begin{equation}\label{wave equation}
\Big[\nabla \times {\mu ^{ - 1}}({\bf r},\omega ) \nabla \times \,-\,\frac{{{\omega ^2}}}{{{c^2}}}\varepsilon ({\bf r},\omega )\Big]{\hat{\bf E}}({\bf r},\omega ) = i\omega {\mu _0}{{\hat{\bf J}}^N}({\bf r},\omega ),
\end{equation}
where $c$ is the speed of light, and $\varepsilon  = 1 + {\chi _e}$ and ${\mu ^{ - 1}} = 1 - {\chi _m}$ are, respectively, the electric permittivity and the inverse magnetic permeability function of the medium.
%The current density operator ${{\bf{\hat J}}^N}({\bf{r'}},\omega )$ is also expressed as a function of the noise polarization and noise magnetization density operators as ${{\bf{\hat J}}^N}({\bf{r'}},\omega ) = - i\omega {{\bf{\hat P}}^N}({\bf{r'}},\omega ) + \vec \nabla  \times {{\bf{\hat M}}^N}({\bf{r'}},\omega )$ \cite{11}. \\

%From now on, for simplicity, we concentrate our attention on
Let us consider a simplified system of two non-relativistic particles of opposite charges ${e_1}=-{e_2}=e$ and masses ${m_1}$ and ${m_2}$ which
form an atomic system, interacting with the electromagnetic field in the presence of a lossy medium.
To separate the degrees of freedom due to the atomic gross motion
from those of the internal motion, it is convenient
to introduce the center-of-mass frame with
new variables,
${\hat{ \bf{R}}} = \sum_{\alpha  = 1}^2 {\frac{{{m_\alpha }{{\hat {\bf{r}}}_\alpha }}}{M}}$ and ${{{\hat {\bf r}}_{rel}}}={{\hat {\bf{r}}}_{1}}-{{\hat {\bf{r}}}_{2}}$ with the corresponding conjugate momenta ${\hat {\bf{\Pi} }}({{\hat {\bf{r}}}_\alpha })={{\hat {\bf{p}}}_1}({{\hat {\bf{r}}}_1})+{{\hat {\bf{p}}}_2}({{\hat {\bf{r}}}_2})
$ and ${\hat {\bf{\pi} }}({{\hat{ \bf{r}}}_\alpha }) = \mu ({{\dot{\hat{ \bf r}}}_1} - {{\dot{\hat {\bf r}}}_2})$, where $M={m_1}+{m_2}$ and $\mu = \frac{{{m_1}{m_2}}}{M}$ are the atomic and reduced mass, respectively.
From the above definition of the new variables and making use of the commutation relations~(\ref{commutation relation of particle}), it immediately follows that
\begin{equation}\label{commutation relation of center-of-mass}
[{\hat R}_i \,,\,{\hat \Pi} _j] = i\hbar {\delta _{ij}},
\end{equation}
\begin{equation}\label{commutation relation of relative coordinat}
[{\hat r}^{rel}_i\,,\,{\hat \pi} _j] = i\hbar {\delta _{ij}}.
\end{equation}
In the following, we rephrase the Hamiltonian~(\ref{Hamiltonian}) in terms of
the center-of-mass dynamical variables. This can be carried out by applying a canonical transformation to Eq.~(\ref{Hamiltonian}) and, when explicitly written, is seen to lead to the usual multipolar Hamiltonian.
The canonical transformation is nothing other than the Power-Zienau-Woolley transformation which is characterized by the unitary operator $\hat U = {e^{i\hat \Lambda }} = \exp\left[ { \frac{i}{\hbar }\int  d^3{\bf r}} \,{{{\hat{\bf P}}}_{at}}({\bf r},t) \cdot {\hat{\bf A}}({\bf r},t) \right]$ where
\begin{equation}\label{atomic polarization}
{{\hat{\bf P}}_{at}}({\bf r},t) = \sum\limits_{\alpha  = 1}^2 {e_\alpha ({{{\hat{\bf r}}}_\alpha } - {\hat{\bf R}})\int_0^1 d\sigma \,\delta \Big( {{\bf r} - {\hat{\bf R}} - \sigma ({{{\hat{\bf r}}}_\alpha } - {\hat{\bf R}})} \Big)}.
\end{equation}
is the atomic polarization field relative to the center-of-mass
coordinate ${\hat {\bf R}}$~\cite{Guilot2002}.
Under this canonical transformation, it is clearly seen that the variables ${{\hat{ \bf{r}}}_\alpha }$, ${\hat{\bf A}}$, and ${\hat{\bf B}}$ are unchanged. In contrast, the canonical momentum ${\hat {\bf p}}_\alpha$ and the transverse part of the electric field ${\hat{\bf E}}^{\bot}$ are modified as follows~\cite{Lembessis1993,Guilot2002}
\begin{eqnarray}\label{definition of p in multipolar coupling}
{\hat{\bf p}}^{'}_{\alpha} &=& {{\hat U}^\dag }{{{\hat{\bf p}}}_\alpha }\hat U \nonumber\\
& = &{{{\hat{\bf p}}}_\alpha } + {e_\alpha }{\hat{\bf A}}({{{\hat{\bf r}}}_\alpha }) + \int d^3{\bf r} \,{{{\hat{\bf \Xi }}}_{at,\alpha} }({\bf r}) \times {\hat{\bf B}}({\bf r}),
\end{eqnarray}
\begin{eqnarray}\label{definition of E in multipolar coupling}
{\hat{\bf E}}^{'\bot} ({\bf r},t) &=& {\hat U^\dag }{{\hat{\bf E}}^ \bot }({\bf r},t)\,\hat U \nonumber\\
& = & {{\hat{\bf E}}^\bot }({\bf r},t)\, - \frac{1}{{{\varepsilon _0}}}{\hat{\bf P}}_{at}^\bot ({\bf r},t),
\end{eqnarray}
where ${\nabla ^\alpha }$ refers to differentiation with respect
to the coordinate ${{\hat {\bf{r}}}_\alpha }$, the subscript $\bot$ indicates
the transverse part of the relevant vector, and ${{\hat{\bf \Xi }}_{at,\alpha} }({\bf{r}})$ is the atomic magnetization field relative to the center-of-mass
coordinate ${\hat {\bf R}}$, which is defined as:
\begin{equation}\label{definition of Xi}
{{\hat{\bf \Xi }}_{at,\alpha} }({\bf{r}}) = \sum_{\beta  = 1}^2 {{e_\beta }\int_0^1 d\sigma \Big( \sigma \delta _{\alpha \beta } - \frac{m_a}{M}[\sigma  - 1] \Big)[{\hat{\bf r}}_\beta }-{\hat{\bf R}}]\delta \Big({\bf r} - {\hat{\bf R}}- \sigma ({\hat{\bf r}}_\alpha - {\hat{\bf R}}) \Big).
\end{equation}

Now, we obtain the multipolar Hamiltonian, $\hat H_{mult}={\hat U^\dag }{\hat H} \,\hat U$, in the electric dipole approximation.
%We denote by $H_{mult}$ the multipolar Hamiltonian given by .
 In order to get the electric dipole approximation of the multipolar Hamiltonian, we expand the Dirac $\delta$ distributions appearing in the atomic polarization vectors ${{\hat{\bf P}}_{at}}$ and ${{\hat{\bf \Xi }}_\alpha }$ in powers of ${{\hat {\bf{r}}}_\alpha }-{\hat {\bf{R}}}$ and keeping only the leading terms.
Using the electric dipole version of Eqs.~(\ref{atomic polarization}) and~(\ref{definition of Xi}), after some lengthy manipulations the multipolar Hamiltonian in the electric dipole approximation is obtained as:
\begin{eqnarray}\label{Hamiltonian in multipolar coupling}
{\hat H}_{mult} &=& \frac{{\hat {\bf{\Pi} }}^2}{2M}+\Big( \frac{{\hat {\bf{p}}}^2}{2\mu }-\frac{e^2}{4\pi \varepsilon _0 {\hat{\bf r}}_{rel}} \Big) + \frac{1}{2}\int d\,^3{\bf r} [\varepsilon _0 {\hat {\bf E}}^{\bot 2}({\bf r},t) + \frac{{\hat {\bf B}}^2({\bf r},t)}{\mu _0}]\nonumber \\
&+& \frac{1}{2}\int d\,^3{\bf r} \int_0^\infty d\omega \Big[ {\dot{\hat {\bf X}}}_\omega ^2({\bf r},t) + {\dot{\hat {\bf Y}}}_\omega ^2({\bf r},t) + {\omega ^2}\Big( {\hat {\bf X}}^{2} _\omega ({\bf r},t) + {\hat {\bf Y}}^{2} _\omega ({\bf r},t) \Big) \Big] \nonumber \\
&-& \int d\,^3{\bf r} \,{\hat {\bf M}}({\bf r},t) \cdot {\hat {\bf B}}({\bf r},t) - \int d\,^3{\bf r} \,{\hat{\bf d}} \cdot {\hat{\bf E}}({\hat{\bf R}})\nonumber \\
&+& \frac{1}{2M}\int d\,^3{\bf r}\, \big({\hat {\bf{\Pi}}} \cdot {\hat {\bf d}} \times {\hat{\bf B}}({\hat{\bf R}}) + {\hat{\bf d}} \times {\hat{\bf B}}({\hat{\bf R}}) \cdot {\hat{\bf{\Pi}}}\big) \,+ \frac{{\hat {\bf d}} \times {\hat {\bf B}}({\bf R})}{8\mu }\nonumber \\
&+& \frac{1}{2\varepsilon _0}\int d\,^3{\bf r} \,{\hat {\bf P}}_{at}^ \bot  + {w_{coul}}\, + (magnetic - dipole\,term),
\end{eqnarray}
where ${\hat{\bf d}} = e{\hat {\bf r}}_{rel}$ is the atomic transition dipole momentum in the laboratory frame with matrix elements ${{\bf{d}}_{kl}}=\langle k|{\hat {\bf{d}}}|l\rangle$,
the ${\hat {\bf \Pi}}$-dependent terms refer to the R\"{o}ntgen interaction which couples the electric moment of the atom to the magnetic field by virtue of the motion of the atom, the magnetic dipole term comes from the product between ${\hat{\bf{\pi}}}$ and ${\hat {\bf{d}}} \times {\hat {\bf{B}}}({\hat {\bf{R}}})$ and would be negligible compared to the electric dipole interaction~\cite{Lembessis1993}, the term $\frac{{[{\hat {\bf{d}}} \times {\hat {\bf{B}}}({\hat {\bf{R}}})]}}{8\mu}$ represents the diamagnetic interaction energy and is small enough
to ignore in our calculations~\cite{Bethe1957}, and the term involving the integral
of the square of the atomic polarization contributes to the Lamb shift and may be omitted
in renormalized energies of the internal motion.

Let us model the atom as a two-level atom with the ground
state $|l\rangle$, the excited state $|u\rangle$ and the energy difference $\hbar{\omega _A}$. With above points in view and introducing the explicit information about the energy levels to Eq.~(\ref{Hamiltonian in multipolar coupling}), we arrive at the following effective Hamiltonian
\begin{eqnarray}\label{effective Hamiltonian}
{\hat H}_{mult}^{\rm eff} &=& \frac{{\hat {\bf{\Pi }}}^2}{2M} + \hbar {\omega _A}{\hat \sigma ^\dag }{\hat \sigma}  + \frac{1}{2}\int d\,^3{\bf r} \Big(\varepsilon _0 {\hat {\bf E}}^{ \bot 2}({\bf r},t) + \frac{{\hat {\bf B}}^2({\bf r},t)}{\mu _0} \Big)\nonumber \\
&+& \frac{1}{2}\int d\,^3{\bf r} \int_0^\infty d\omega \Big[ {\dot{\hat {\bf X}}}_\omega ^2({\bf r},t) + {\dot{\hat {\bf Y}}}_\omega ^2({\bf r},t) + {\omega ^2}\Big( {\hat {\bf X}}^{2} _\omega ({\bf r},t) + {\hat {\bf Y}}^{2} _\omega ({\bf r},t) \Big) \Big] \nonumber \\
&-& \int d\,^3{\bf r} \,{\hat {\bf M}}({\bf r},t) \cdot {\hat {\bf B}}({\bf r},t) - \int d\,^3{\bf r} \,{\hat {\bf d}} \cdot {\hat{\bf E}}({\hat {\bf R}})\nonumber \\
&+& \frac{1}{2M}\int d\,^3{\bf r}\, \big({\hat{\bf{\Pi}}} \cdot {\hat{\bf d}} \times {\hat{\bf B}}({\hat{\bf R}}) + {\hat{\bf d}} \times {\hat{\bf B}}({\hat{\bf R}}) \cdot {\hat{\bf{\Pi}}}\big) ,
\end{eqnarray}
where $\hat \sigma$ and ${\hat \sigma ^\dag}$ are the Pauli lowering and raising operators of the two-level atom.
%and the last term in equation (\ref{effective Hamiltonian}) represents the R\"{o}ntgen interaction term that is introduced into Hamiltonian as a result of the movement of the center of mass of the atomic system.
%

To facilitate calculations, we introduce the bosonic operators ${\hat{\bf f}_e}({\bf r},\omega ,t) = \frac{1}{\sqrt {2\hbar \omega}}[ - i\omega {{\hat {\bf X}}_\omega }({\bf r},t) + {{\hat {\bf Q}}_\omega }({\bf r},t)]$ and ${\hat{\bf f}_m}({\bf r},\omega ,t) = \frac{1}{\sqrt {2\hbar \omega}}[\omega {{\hat{\bf Y}}_\omega }({\bf r},t) + i{{\hat{\bf \Pi }}_\omega }({\bf r},t)]$ which play the roll of the collective excitations of the electromagnetic field and the absorbing medium. By lengthy but straightforward calculations, one can show that the Hamiltonian of the electromagnetic field, the medium and the interactions between them, i.e. the last term in the first line, the second line and the first term in the third line of Eq.~(\ref{effective Hamiltonian}), are simplified as~\cite{Amoghorban2010,Huttner1992,Philbin2010}
\begin{equation}\label{Hamiltonian of matter}
{\hat H_F} = \sum\limits_{\lambda  = e,m} \int d\omega \int d\,^3{\bf r}\, \hbar \omega \,{{\hat{\bf f}}_\lambda }^{\dag} ({\bf r},\omega ,t) \cdot {{\hat{\bf f}}_\lambda }({\bf r},\omega ,t),
\end{equation}
where ${\hat{\bf f}_e}({\bf r},\omega ,t)$ and ${\hat{\bf f}_m}({\bf r},\omega ,t)$ associated with the electric and magnetic excitations.
By applying the commutation relations~(\ref{commutation relation of X}) and~(\ref{commutation relation of Y}), it can be shown that the bosonic operators satisfy the usual commutation relations
\begin{equation}\label{commutation relations for bosonic operators}
[{\hat f_{\lambda ,j}}({\bf r},\omega ,t)\,,\,{\hat f^\dag }_{\lambda ',j'}({\bf r'},\omega ',t)] = {\delta _{jj'}}\delta ({\bf r} - {\bf r'})\delta (\omega  - \omega ').
\end{equation}
With the help of these bosonic operators and Eqs.~(\ref{wave equation}),~(\ref{noise polarization}) and~(\ref{noise magnetization}), the explicit form of the electric field operator is obtained in terms of the electromagnetic Green tensor of the system as follows:
%Using these operators, and substituting them in the wave equations , the electric field operator is written as follows \cite{2}
%
\begin{eqnarray}\label{electric field}
{\hat{\bf E}}({\bf r},t) &=& i\sqrt {\frac{{\hbar {\mu _0}}}{\pi}} \int d\,^3{\bf r'}\,\int_0^\infty  d\omega \,\omega \frac{\omega }{c}\sqrt{{\rm Im} \varepsilon ({\bf r'},\omega )}{\bfsfG}({\bf r},{\bf r'},\omega ) \cdot {\hat{\bf f}}_e({\bf r'},\omega )\nonumber \\
&+& \sqrt {\frac{{\rm Im}\mu ({\bf r'},\omega )}{|\mu ({\bf r'},\omega )|^2}} \Big[ \nabla' \times {\bfsfG}({\bf r'},{\bf r},\omega )\Big]^T \cdot {\hat{\bf f}}_m({\bf r'},\omega ) + H \cdot C\cdot,
\end{eqnarray}
where ${\bfsfG}({\bf r},{\bf r'},\omega )$ is the unique solution to the Helmholtz equation
\begin{equation}\label{Helmholtz equation}
\Big[ \nabla \times \mu ^{ - 1}({\bf r},\omega ) \nabla \times\,- \,\frac{\omega ^2}{c^2}\varepsilon ({\bf r},\omega )\Big] {\bfsfG}({\bf r},{\bf r'},\omega ) = \delta ({\bf r} - {\bf r'}){\bar{\bar I}}\,,
\end{equation}
together with the boundary condition $\bfsfG ({\bf r},{\bf r}',\omega)\rightarrow 0$ for $|{\bf r}-{\bf r}'| \rightarrow \infty$, satisfies the integral relation
\begin{eqnarray}\label{integral relation of Green tensor}
&&\int d\,^3{\bf s}\, \Big(- {\rm Im} [{\mu ^{ - 1}}({\bf s},\omega )][\nabla _{\bf s} \times {\bfsfG}({\bf s},{\bf r},\omega )]^T \cdot [{ \nabla _{\bf s}} \times {\bfsfG^*}({\bf s},{\bf r'},\omega )]\nonumber \\
&&+ \frac{\omega^2}{c^2}{\rm Im} [\varepsilon ({\bf s},\omega )]{\bfsfG}({\bf r},{\bf s},\omega) \cdot {{\bfsfG^*}({\bf s},{\bf r'},\omega )} \Big)= {\rm Im}[{\bfsfG}({\bf r},{\bf r'},\omega )].
\end{eqnarray}
All the information about the geometry and topology of the environmental media are contained in the Green tensor of the system ${\bfsfG}$ via the electric permittivity function $\varepsilon$ and the magnetic permeability function $\mu$.
Now, the set of equations~(\ref{effective Hamiltonian}), (\ref{Hamiltonian of matter}) and (\ref{electric field}) together with the commutation relations~(\ref{commutation relation of center-of-mass}), (\ref{commutation relation of relative coordinat}) and (\ref{commutation relations for bosonic operators}) provide
the canonical quantization of the electromagnetic field interacting with moving charge particles in the presence of an isotropic, inhomogeneous and absorbing magnetodielectric medium.
As we expected, these equations are the same as obtained from the phenomenological method~\cite{Buhmann2012a,Buhmann2012b}.
In recent years, based on the phenomenological approach, the quantum description of moving atoms at non-relativistic speeds has been studied extensively~\cite{Scheel2009,Buhmann2012a,Buhmann2012b}.
In the next section, as an application of our formalism and a main purpose of this article pointed out earlier, we treat the general case that the atomic system is in uniform motion with relativistic speeds in the presence of material medium.
%
%%%%%%%%%%%%%%%%%%%%%%%%%%%%%%%%%%%%%%%%%%%%%%%%%%%%%%%%%%%%%%%%%%
\section{Relativistic dynamics}\label{Sec:Relativistic dynamics}
In this section, we study the dynamics of a moving two-level atom interacting with the electromagnetic field near absorbing media at the relativistic regime.
The starting point is the total Lagrangian~(\ref{Lagrangian}), with the only difference is that the particle motion is treated relativistically. Because the medium is at rest, therefore, there is no need to write the Lagrangian density associated with the medium in a covariant form.
Given that the internal dynamics of the atom are not affected by relativistic considerations other than through ${\hat {\bf R}}$~\cite{Boussiakou2002}, we therefore follow the approach presented in~\cite{Amoghorban2010,Amoghorban2014} and describe the center-of-mass dynamical variables by the Dirac field. To do so, we start from the total Lagrangian~(\ref{Lagrangian}), and rewrite the Lagrangian associated with the kinetic energy of the charged particles and their interactions with the electromagnetic field in term of the center-of-mass coordinate ${\hat {\bf R}}$ and the relative coordinate ${\hat {\bf r}}_{rel}$.
%Since, the medium is at rest and the atom is in motion, we do not need to write the medium part of the Lagrangian in a covariant form.
With these notes, we can replace the relevant terms associated to the center-of-mass coordinate by
\begin{eqnarray}\label{relativistic Lagrangian}
{L_D} = \frac{i\hbar c}{2}\int d\,^3{\bf r} \left\{ {\sum_{\mu = 1}^4 {\left( {\bar \psi ({\bf r},t){\gamma ^\mu }\frac{{\partial \psi ({\bf r},t)}}{{\partial {{\rm r}^\mu}}} - \frac{{\partial \bar \psi ({\bf r},t)}}{{\partial {{\rm r}^\mu }}}{\gamma ^\mu }\psi ({\bf r},t)} \right)}} \, - M{c^2}\bar \psi \psi\right\},\ \nonumber\hspace{-1cm}\\
%&+& e_T\int d^3{\bf r} \left\{ {\sum\limits_{j = 1}^3 {\left( {c\bar \psi ({\bf r},t){\gamma ^j}\psi ({\bf r},t){{\bf A}^j}({\bf r},t)} \right)}  - \bar \psi %({\bf r},t){\gamma ^0}\psi ({\bf r},t)\varphi ({\bf r},t)} \right\}.
\end{eqnarray}
and write the external current and charge densities, respectively, as
$ {\bf J}({\bf r},t) =ec  {\psi}^\dag ({\bf r},t) {\boldsymbol \alpha} \psi ({\bf r},t)  $
and
$ \rho({\bf r},t)= e {\psi}^\dag \psi ({\bf r},t),$
where ${\gamma^\mu }$ are the Dirac matrices with ${\gamma^4} = \beta$, ${\gamma ^j} = \beta {\alpha _j}$, and ${\bar \psi}  = {\psi ^\dag }\beta$.
In the standard representation, the ${\gamma}$ matrices are expressed in terms of the $2\times 2$
Pauli matrices $\sigma_j\,(j=1,2,3)$ and the identity matrix $\mathbb{I}$, as
\begin{eqnarray}\label{matrices}
\beta  = \left( {\begin{array}{*{20}{c}}
\mathbb{I}&0\\
0&{ - \mathbb{I}}
\end{array}} \right),\,\,\,\,\,\,{\alpha _j} = \left( {\begin{array}{*{20}{c}}
0&{{\sigma _j}}\\
{{\sigma _j}}&0
\end{array}} \right)\,,
\end{eqnarray}
The canonical momentum associated to $\psi$ is given by ${{i\hbar {\psi ^\dag }}}/{2} = \frac{\partial L}{{\partial \dot \psi ({\bf r},t)}}$. Now, we impose the canonical anti-commutation relation for the Dirac field as
\begin{equation}\label{first anticommutation relation}
\big\{ {\psi _\alpha ({\bf r},t)\,,\,\psi _\beta ^\dag ({\bf r'},t)} \big\} = \delta _{\alpha \beta}\,\delta ({\bf r} - {\bf r'}),
\end{equation}
\begin{equation}\label{second anticommutation relation}
\big\{ {\psi _\alpha ({\bf r},t)\,,\, \psi _\beta ({\bf r'},t)} \big\} = 0.
\end{equation}
%
%The Hamiltonian of the system is obtained by substituting Eq.~(\ref{relativistic Lagrangian}) into~(\ref{Lagrangian}) and making use of Eq.~(\ref{conjugate momentum of field})-(\ref{conjugate momentum of Y}) and the canonical momentum $ {\psi ^\dag }$. We find that
%
%\begin{eqnarray}\label{relativistic Hamiltonian}
%H &=& \int d\,^3{\bf r} (- i\hbar c \psi ^\dag ({\bf r},t){\bf \alpha} \cdot \nabla \psi ({\bf r},t) + Mc^2 \psi ^\dag({\bf r},t)\beta \psi ({\bf r},t)\nonumber \\
%&+& \hbar\omega _0 {{\bf \pi}^\dag}{\bf \pi} + \frac{1}{2}\int d\,^3{\bf r}[\varepsilon _0 {{\bf E}^{\bot 2}}({\bf r},t) + \frac{{\bf B}^2 ({\bf r},t)}{\mu _0}]\nonumber \\
%&+& \frac{1}{2}\int d\,^3{\bf r} \int_0^\infty d\omega \,[{\bf Q}_\omega ^2({\bf r},t) +  \omega ^2 {\bf X}_\omega ^2 ({\bf r},t)]\nonumber \\
%&+& \frac{1}{2} \int d\,^3 {\bf r} \int_0^\infty d\omega \,[{\bf \Pi}_\omega ^2({\bf r},t) + \omega ^2 {\bf Y}_\omega ^2({\bf r},t)]\nonumber \\
%&-& \int d\,^3 {\bf r} [{\bf M}({\bf r},t) \cdot {\bf B}({\bf r},t) + {\bf \dot P}({\bf r},t) \cdot {\bf A}({\bf r},t)]\nonumber \\
%&-& \frac{1}{2}\int d\,^3{\bf r} \int_0^\infty d\omega \, g_e ({\bf r},\omega ) {\bf A}^2({\bf r},t).
%\end{eqnarray}
%
It is convenient to expand the Dirac field $\psi ({\bf r},t)$ in terms of the eigenfunctions of Dirac equation in the absence of electromagnetic field
\begin{equation}\label{wave function}
\psi ({\bf r},t) = \frac{1}{(2\pi)^{3/2}}\sum_{\mu = 1}^4 \int d\,^3 {\bf q} \, \hat c_\mu ({\bf q},t) \psi _\mu ({\bf q}),
\end{equation}
where $\psi_\mu ({\bf q}) = u_\mu ({\bf q})\, e^{i{\bf q} \cdot {\bf r}}$ with $ u_\mu ({\bf q})$ are the four-component spinors and ${\hat c}_\mu ({\bf q},t)$ is the atomic annihilation operator~\cite{Bogoliubov1980}. With the help of the above expansion and by substituting Eq.~(\ref{relativistic Lagrangian}) into~(\ref{Lagrangian}) and making use of Eqs.~(\ref{conjugate momentum of field})-(\ref{conjugate momentum of Y}) and the canonical momentum $ {\psi ^\dag }$, we obtain the effective multipolar Hamiltonian as
%
%Substituting Eq.~(\ref{wave function}) into~(\ref{relativistic Hamiltonian}) and using Eqs.~(\ref{conjugate momentum of X}) and (\ref{conjugate momentum of Y}), the effective multipolar Hamiltonian is obtained as
%
\begin{equation}\label{final Hamiltonian}
{\hat H}_{mult} = \hat H_A + \hat H_F + \hat H_{int},
\end{equation}
where $ \hat H_F$ is the Hamiltonian given by Eq.~(\ref{Hamiltonian of matter}) and $\hat H_A = \frac{1}{2}\frac{\hbar \omega _A}{\gamma} {\hat\sigma}^{\dag} {\hat\sigma}  + \sum_{\mu=1}^4\int d\,^3 {\bf q}\,E_{\bf q,\mu}\, \hat c^\dag_\mu ({\bf q},t) \hat c_\mu({\bf q},t)$ denotes
the Hamiltonian of the moving atom, wherein $\gamma=\sqrt{1 - {\dot{\bf  R}^2}/{c^2}}$ with $\dot{\bf  R}$ is the center-of-mass velocity of the atom. Here, the first term in the right-hand side of $\hat H_A$ corresponds to the internal dynamics of the atom in the laboratory frame, while the second term is related to the center-of-mass motion of the atom with the momentum $\hbar {\bf q}$ and the energy $E_{\bf q,\mu}=\pm\sqrt {\hbar ^2 c^2 q^2+M^2 c^4  }$, in which the plus(minus) sign is to be taken for $\mu=1,2$($\mu=3,4$). The unperturbed Hamiltonian $\hat H_A + \hat H_F $ has the eigenstate
$|atom + rad\rangle  = |atom\rangle |rad\rangle$ which are the direct product of the eigenstates of $\hat H_A  $ and $ \hat H_F $.

The interaction Hamiltonian $\hat H_{int}$ in Eq.~(\ref{final Hamiltonian}) composed of two parts: the interaction between the electric dipole and the electric and magnetic field, and the R\"{o}ntgen interaction. The interaction Hamiltonian in the Fourier space is given by
\begin{eqnarray}\label{relativistic interaction Hamiltonian}
{\hat H}_{int}^{\rm eff}&=& - \frac{i\mu _0}{(2\pi )^3} \hat \sigma ^\dag {\hat{\bf{d}}_A} \cdot \sum_{\mu,\mu'=1}^4 \int_0^\infty  d\omega \omega \int d\,^3 {\bf q}\int d\,^3{\bf k}\int d\,^3 {{\bf k}_1} \int d\,^3 {{\bf k}_2}\,\nonumber \\
&& \times{\bfsfG}({{\bf k}_2},{\bf k},\omega)\,\Big[\omega g_e({\bf k} - {{\bf k}_1},\omega) \cdot \hat {\bf f}_e ({{\bf k}_1},\omega) + i{\bf k} g_m({\bf k} - {{\bf k}_1},\omega) \cdot \hat {\bf f}_m({{\bf k}_1},\omega)\Big]\nonumber \\
&&\times u_\mu^\dag({\bf q})  u_{\mu'}({\bf q}-{{\bf k}_2}) \hat c_\mu^\dag ({\bf q},t)\,\hat c_{\mu'}({\bf q}-{{\bf k}_2},\omega) + H\cdot C\cdot \nonumber \\
&-& \frac{i\mu _0}{(2\pi )^3} \hat \sigma ^\dag {{\bf d}_A} \cdot \sum_{\mu,\mu'=1}^4 \int_0^\infty d\omega \int d\,^3{\bf q}\int d\,^3 {\bf k}\int d\,^3 {{\bf k}_1} \int d\,^3{{\bf k}_2}\nonumber \\
&&\times{\dot {\bf R}} \times {\bf k}_2 \times {\bfsfG}({\bf k}_2,{\bf k},\omega )\Big[\omega g_e({\bf k} - {\bf k}_1,\omega)\cdot \hat {\bf f}_e({\bf k}_1,\omega)\nonumber \\
&&+ i{\bf k} g_m({\bf k} - {\bf k}_1,\omega )\cdot \hat {\bf f}_m({\bf k}_1,\omega)\Big]\nonumber \\
&&\times u_\mu^\dag({\bf q})  u_{\mu'}({\bf q}-{{\bf k}_2}) \hat c_\mu^\dag ({\bf q},t)\,\hat c_{\mu'}({\bf q}-{{\bf k}_2},\omega)+ H\cdot C\cdot ,
\end{eqnarray}
where ${{{\bf{d}}}_A} = \langle l|{{\hat{\bf d}}}|u\rangle  = \langle u|{{\hat{\bf d}}}|l\rangle$.
%is the transition dipole matrix element in the rest frame of the atom.
%
%%%%%%%%%%%%%%%%%%%%%%%%%%%%%%%%%%%%%%%%%%%%%%%%%%%%%%%%%%%%%%%%%%%%%%%%%%%%%%%%%%%%%%%%%%5
%
\section{Radiative properties of a moving atom}\label{Sec:Radiative properties of a moving atom}
In this section, to describe the radiative dynamics of the moving
atom, we study the time evolution of the system state vector and find their probability amplitudes by using the Markov approximation.
For a single-quantum excitation, the time-dependent state vector of the whole system can be written as
\begin{eqnarray}\label{state vector}
|\psi ({\bf q},t)\rangle  &=& \sum_{\mu=1}^2\int d\,^3{\bf q}\, C_{u,\mu}({\bf q},t) e ^{- i\tilde \omega _A t}|{\bf q}  \rangle_\mu  |u\rangle |\{0\} \rangle \nonumber \\
&+&\sum_{\lambda = e,m} \sum_{\mu=1}^2\int d\,^3{\bf q} \int d\,^3{\bf k} \int d\,^3 {\bf k'} \int d\omega e ^{- i(\omega + \omega _B)t}\nonumber \\
&& \times C_{\lambda l,\mu} ({\bf q},{\bf k},{\bf k'},\omega ,t)|{\bf q}-{\bf k'}\rangle_\mu |l\rangle  |\{{\bf 1}_\lambda ({\bf k'},\omega)\} \rangle .
\end{eqnarray}
where the atomic state vectors  $|{\bf q}\rangle_\mu$ and $|{\bf q} - {\bf k'}\rangle_\mu$ refer to the momentum of the atom before and after the emission of the photon, $| \{0\} \rangle$ is the vacuum state of the field, $|\{{\bf 1}_\lambda ({\bf k},\omega)\} \rangle$ is the excited state of the field with ${\hat {\bf f}}^\dag_\lambda ({{\bf k}},\omega)|\{0\} \rangle =|\{{\bf 1}_\lambda ({\bf k},\omega)\} \rangle$, and $ C_{u,\mu}({\bf q},t)$ and $C_{\lambda l,\mu}({\bf q},{\bf k},{\bf k'},\omega ,t)$ are the respective probability amplitudes of the excited and ground states of the system,
%Also, $|\{0\} \rangle$ represents the ground state of the system composed of electromagnetic field and matter. Here, the coefficients $ C_{u,{\bf q}}(t)$ and $C_{\lambda l,{\bf q}}({\bf k},{\bf k'},\omega ,t)$ are the probability amplitudes of finding the system in the initial state $|{\bf q}\rangle |\{0\} \rangle |u\rangle$ and the final state $|{\bf q} - {\bf k'}\rangle |\{1_\lambda ({\bf k'},\omega) \} \rangle |l\rangle$, respectively.
the subscript $\lambda = e,\,m$ indicates the electric and magnetic excitations of the medium, and
the frequencies $\tilde \omega _A$ and $\omega_B$ are, respectively, defined as $\tilde \omega _A =\frac{\omega_A}{\gamma} + \frac{|E_{\bf q}|}{\hbar}-\delta \omega$ and $\omega_B = \frac{|E_{\bf q - k'}|}{\hbar}$ with $\delta \omega$ denoting the Lamb shift.
In this case the dynamics is described by the time-dependent Schr\"{o}dinger equation $i\hbar \partial _t |\psi (t)\rangle  = \hat H |\psi (t)\rangle$. We insert the state vector~(\ref{state vector}) into the Schr\"{o}dinger equation, and obtain the following equations of motion:
\begin{eqnarray}\label{time evaluation of C_q}
\dot C_{u,\mu}({\bf q},t) &=& - i\delta \omega C_{u,\mu}({\bf q},t) \nonumber \\
& -& \frac{1}{(2\pi )^3 c\sqrt {\pi \varepsilon _0 \hbar}}\sum_{\mu'=1}^2 \int_0^\infty {d\omega} \,e ^{- i(\omega  + \omega _B - \tilde \omega _A)t} \int  d\,^3 {\bf k} \int d\,^3{\bf k'} \int d\,^3 {\bf k}_1\nonumber\\
&&\times {\hat{\bf d}}_A \cdot\Bigg\{ \Big[\frac{\omega ^2}{c} g_e ({\bf k}_1 - {\bf k},\omega ){\bfsfG}({\bf k'},{\bf k}_1,\omega) C_{el,\mu'}({\bf q},{\bf k},{\bf k'},\omega ,t)\nonumber \\
&&+ i \omega  g_m({\bf k}_1 - {\bf k},\omega ){\bfsfG}({\bf k'},{\bf k}_1,\omega) \times {\bf k }_1 C_{ml,\mu'}({\bf q},{\bf k},{\bf k'},\omega ,t) \Big] \\
&&+ {\dot {\bf R}} \times {\bf k'} \times \Big[\frac{\omega }{c} g_e({\bf k}_1 - {\bf k},\omega){\bfsfG}({\bf k'}, {\bf k}_1,\omega) C_{el,\mu'}({\bf q},{\bf k},{\bf k'},\omega ,t)\nonumber \\
&&+ i g_m({\bf k}_1 - {\bf k},\omega){\bfsfG}({\bf k'}, {\bf k}_1,\omega) \times {\bf k}_1 C_{ml,\mu'}({\bf q},{\bf k},{\bf k'},\omega ,t) \Big] \Bigg\}\nonumber\\
&& \times u_\mu^\dag({\bf q})  u_{\mu'}({\bf q}-{{\bf k}'}) \nonumber ,
\end{eqnarray}
\begin{eqnarray}\label{time evaluation of C_e}
\dot C_{el,\mu}({\bf q},{\bf k},{\bf k'},\omega ,t) &=&\frac{1}{(2\pi )^3 \sqrt {\pi \varepsilon_0 \hbar}}\sum_{\mu'=1}^2\int d\,^3 {\bf k}_1 \frac{\omega }{c^2} e^{- i(\tilde \omega _A -\omega - \omega _B )t} C_{u,\mu'}({\bf q},t) \nonumber \\
&& \times g_e^* ({\bf k}_1 - {\bf k},\omega ) u_\mu^\dag({\bf q}-{{\bf k}'})  u_{\mu'}({\bf q}) \Big\{ \omega {\bfsfG^*}(-{\bf k}_1 , -{\bf k}',\omega ) \nonumber \\
&&+ {\bfsfG^*} (- {\bf k}_1 , -{\bf k}',\omega) \times {\bf k'} \times {\dot {\bf R}} \Big\} \cdot {\hat{\bf d}}_A ,
\end{eqnarray}
\begin{eqnarray}\label{time evaluation of C_m}
\dot C_{ml,\mu}({\bf q},{\bf k},{\bf k'},\omega ,t) &=& \frac{1}{(2\pi )^3 \sqrt {\pi \varepsilon _0 \hbar}}\sum_{\mu'=1}^2\int d\,^3 {\bf k}_1 \frac{1}{c} e^{- i(\tilde \omega _A - \omega  - \omega _B)t} C_{u,\mu'}({\bf q},t) \nonumber \\
&&\times g_m^* ({\bf k}_1 - {\bf k},\omega) u_\mu^\dag({\bf q}-{{\bf k}'})  u_{\mu'}({\bf q})\nonumber\\
&&\times\Big\{ i \omega {\bf k}_1 \times {\bfsfG^*} (-{\bf k}_1 , -{\bf k}',\omega) \nonumber \\
&&+ i\,{\bf k}_1 \times {\bfsfG^*} (-{\bf k}_1 , -{\bf k}',\omega) \times {\bf k'} \times {\dot {\bf R}} \Big\}  \cdot {\hat{\bf d}}_A.
\end{eqnarray}
%after lengthly calculations
Suppose that the system is prepared in a state described by $|{\bf q}\rangle_\mu |u\rangle  |\{0\} \rangle $.
%
%Since the main purpose of this paper is to obtain the spontaneous emission rate of a two-level atom, we consider the initial state of the system as $|{\bf q}\rangle |\{0\} \rangle |u\rangle$.
Therefore, at the initial time $t=0$, we have $ C_{u,\mu}({\bf q},0) = 1$ (for $\mu=1$ or $\mu=2$) and $C_{\lambda l,\mu}({\bf q},{\bf k},{\bf k'},\omega ,0) = 0. $
At any time $t>0$, we can calculate the amplitude probability that the moving atom has emitted a photon
%due to the coupling the atom in its excited state couples with the the electromagnetic field and the medium.
by substituting the solution of Eqs.~(\ref{time evaluation of C_e}) and~(\ref{time evaluation of C_m}) into~(\ref{time evaluation of C_q}) and using the integral relation~(\ref{integral relation of Green tensor}) in the Fourier space.  Then, after a lengthy calculation, we obtain
\begin{eqnarray}\label{final time evaluation of C_q}
\dot C_{u}({\bf q},t) &=& - i\delta \omega \,C_{u}({\bf q},t)\nonumber \\
&+& \sum_{\mu,\mu'=1}^2 \int d\,^3{\bf k'} \, u_{\mu}^\dag({\bf q}) u_{\mu'}({\bf q}-{{\bf k}'}) u^\dag_{\mu'}({\bf q}-{{\bf k}'}) u_{\mu}({\bf q})\nonumber\\
&&\times\int_0^t  dt'\,\mathbb{K}({{\bf k}'},t - t') \, C_{u}({\bf q},t'),
\end{eqnarray}
where we have used $ \sum_\mu C_{u,\mu}({\bf q},t)  u_{\mu}({\bf q})= C_{u}({\bf q},t)$. Here, the kernel function $\mathbb{K}({{\bf k}'},t - t')$ can be written in terms of the imaginary part of the Green tensor of the system as
% $  u_\mu^\dag({\bf q}) u_{\mu''}({\bf q})=\delta_{\mu \mu''}$
%
\begin{eqnarray}\label{kernel}
\mathbb{K}({{\bf k}'},t - t') &=& - \frac{1}{(2\pi )^3 (\hbar\pi \varepsilon _0)}\int_0^\infty  \frac{d\omega}{c^2} e^{- i(\tilde \omega _A -\omega -\omega_B)(t' - t)}\nonumber \\
&\times &{\hat{\bf d}}_A \cdot \Big\{ \omega ^2 {\rm Im}{\bfsfG}({\bf k'},\omega) + \omega {\dot {\bf R}} \times {\bf k'}\times {\rm Im}{\bfsfG}({\bf k'},\omega) \\
&+& \omega {\rm Im}{\bfsfG}({\bf k'},\omega) \times {\bf k'} \times {\dot {\bf R}}+{\dot {\bf R}} \times {\bf k'} \times {\rm Im}{\bfsfG}({\bf k'},\omega) \times {\bf k'} \times {\dot {\bf R}} \Big\}  \cdot {{\hat{\bf d}}_A}, \nonumber
\end{eqnarray}
in which we have defined ${\bfsfG}({\bf k'},\omega)={\bfsfG}({\bf k'},{\bf k'},\omega)$.
To make further progress, in first step, it is necessary to compute $\sum_{\mu,\mu'=1}^2  u_{\mu}^\dag({\bf q}) u_{\mu'}({\bf q}-{{\bf k}'}) u^\dag_{\mu'}({\bf q}-{{\bf k}'}) u_{\mu}({\bf q})$. To do so, we can extend the sums over $\mu$ and $\mu'$ to include all four values. We can do this by using
\begin{eqnarray}
\frac{c\boldsymbol{\alpha}\cdot {\bf q}+\beta Mc^2 + |E_{\bf q}|}{2|E_{\bf q}|}u_\mu({\bf q})= \left\{ {\begin{array}{*{20}{c}}
u_\mu({\bf q}) \,\,\,\,\,\,\,\,for \,\,\mu=1,2\\
0 \,\,\,\,\,\,\,\,\,\,\,\,\,\,\,\,\,\,for \,\,\mu=3,4
\end{array}}\right.
\end{eqnarray}
and a similar relation for $ u_{\mu'}({\bf q}-{{\bf k}'})$. Now, we consider the completeness relation $\sum_{\mu=1}^4  u_\mu({\bf q}-{\bf k}')  u^\dag_\mu({\bf q}-{\bf k}')=\mathbb{I}$, therefore, we arrive at
\begin{eqnarray}\label{mu identity}
&&\sum_{\mu,\mu'=1}^2  u_{\mu}^\dag({\bf q}) u_{\mu'}({\bf q}-{{\bf k}'}) u^\dag_{\mu'}({\bf q}-{{\bf k}'}) u_{\mu}({\bf q})=\frac{1}{4|E_{\bf q}||E_{{\bf q}-{{\bf k}'}}|}\nonumber\\
&&\times{\rm Tr} \big[\big(c\boldsymbol{\alpha}\cdot ({{\bf q}-{{\bf k}'}})+\beta Mc^2 + |E_{{\bf q}-{{\bf k}'}}|\big) \big(c\boldsymbol{\alpha}\cdot {\bf q}+\beta Mc^2 + |E_{\bf q}| \big)\big].
%=\frac{1}{2|E_{{\bf q}-{{\bf k}'}}|}{\rm Tr} \big[c\boldsymbol{\alpha}\cdot ({{\bf q}-{{\bf k}'}})+\beta Mc^2 + |E_{{\bf q}-{{\bf k}'}}|\big].
\end{eqnarray}
The above trace can be evaluated using  the identity ${\rm Tr[(\boldsymbol{\alpha}\cdot {\bf a})(\boldsymbol{\alpha}\cdot {\bf b})]}=4{\bf a}\cdot {\bf b}$ and
the fact that the trace of a product of any odd number of the
matrices $\sigma_x$, $\sigma_y$, $\sigma_z$ and $\beta$ is zero. Therefore, Eq.~(\ref{mu identity}) becomes
\begin{equation}
1+\sqrt{(1-v_1^2/c^2)(1-v_2^2/c^2)}+\frac{{\bf v_1} \cdot {\bf v_2}}{c^2},
\end{equation}
where ${\bf v_1}$ and ${\bf v_2}$ are the velocities before and after the emission of the photon. If the momentum of the photon is negligible compared to the center-of-mass momentum of the atom, then ${\bf v_1}\simeq{\bf v_2}$ and the above relation become equal to two. This is true in both the classical limit and the extreme relativistic limit.

In the next step, to obtain an analytic solution for the probability amplitude~(\ref{final time evaluation of C_q}), we restrict our attention to the case that the atom-field system coupled weakly. We use the Markov approximation and replace the coefficient $C_{u}({\bf q},t')$ by $C_{u}({\bf q},t)$, and let the
upper integration limit in the above equation tend to infinity, then approximate the time integral $\int_0^\infty dt' e^{- i(\tilde\omega _A -\omega -\omega_B)(t' - t)} \approx \zeta (\omega  + \omega _B -\tilde\omega _A)$ by the zeta function $\zeta (x) = \pi \delta (x) + iP({1/x})$, in which $P$ stands for Cauchy principal value.
% Let us make some comments on the validity of this approach.
%

Some remarks on the Markov approximation are now in order. First of all, it is interesting to note that there have been various theoretical studies and debates on the problem of the quantum friction force when an atom moves at constant velocity near a macroscopic body~\cite{Pendry1997,Volokitin2006,Philbin2009,Pendry2010,Barton2010a,Barton2010b,Dedkov2002,Golyk2013,Pieplow2013,Intravaia2015,Intravaia2016,Klatt2017}. The result is a velocity dependent force causes the atom to decelerate, but leads to contradictory predictions concerning its dependence on velocity and with atom-surface separation.
%quite contradictory results  there is no way to decide which one is correct
Unfortunately, due to the short range and extremely small magnitude of this force, there is no experimental confirmation yet with presently available technology to judge which of theoretical results is correct. Therefore, we cannot decide whether any of these results is flawed.
To clarify the origins of these disagreement between quantum friction calculations, Intravaia \textit{et al.}~\cite{Intravaia2015,Intravaia2016} showed that
the Markov approximation fails to provide reliable predictions for both in equilibrium and out of equilibrium quantum friction.
With this in view,  Klatt \textit{et al.}~\cite{Klatt2017} based on Markovian quantum master equations and time-dependent perturbation theory have
obtained interesting results concerning the quantum frictional force and the dynamical corrections to the level shifts and decay rates. It is observed that the two approaches agree for the level shifts, the decay rates, and the velocity and distance dependency of the quantum frictional force, if considering the second order in the atom-field coupling. In contrast, both approaches for the quantum friction force differ to the terms of fourth order in coupling~\cite{Klatt2017}.
%
%Whereas, both approaches agree regarding the leading-order dynamical corrections to level shifts and decay rates which is to be studied here.

%give the same contribution to the first line
With the above background and employing the above approximations, we arrive at the familiar result $C_{u,\mu}({\bf q},t) = \exp (- \frac{1}{2}\Gamma  + i\delta \omega)t$ ($\mu=1 $ or  $\mu=2 $), where $\Gamma$ and $\delta$ are, respectively, the decay rate of the excited atom and the Lamb shift in the laboratory frame. These parameters are defined in terms of the Green tensor of the system as:
\begin{eqnarray}\label{relativistic decay rate}
\Gamma  &=& \frac{2}{(2\pi )^3 \hbar\varepsilon _0 c^2}\int d\,^3 {\bf k'} \int d\omega \,\delta (\omega  -(\frac{\omega _0}{\gamma} - \delta \omega) -{\dot {\bf R}} \cdot {\bf k'})\nonumber \\
&&\Big[ {\hat{\bf d}}_A  \cdot \Big\{ \omega ^2 {\rm Im}{\bfsfG}({\bf k'},\omega)+\omega {\dot {\bf R}} \times {\bf k'} \times {\rm Im}{\bfsfG}({\bf k'},\omega)\nonumber \\
&&+ \omega{\rm Im}{\bfsfG}({\bf k'},\omega) \times {\bf k'} \times ({\dot {\bf R}} -\frac{\hbar {\bf k'}}{\gamma M})\nonumber \\
&&+ {\dot {\bf R}} \times {\bf k'} \times {\rm Im}{\bfsfG}({\bf k'},\omega) \times {\bf k'} \times ({\dot {\bf R}} -\frac{\hbar {\bf k'}}{\gamma M}) \Big\}  \cdot {\hat{\bf d}}_A \Big],
\end{eqnarray}
\begin{eqnarray}\label{relativistic Lamb shift}
\delta \omega  &=& \frac{1}{(2\pi )^3 \hbar\varepsilon _0 \pi c^2} \int d\,^3 {\bf k'} \, P\int \frac{d\omega}{\omega  -(\frac{\omega _0}{\gamma} -\delta \omega) -{\dot {\bf R}} \cdot {\bf k'}}\nonumber \\
&&\Big[ {\hat{\bf d}}_A  \cdot \Big\{ \omega ^2 {\rm Im}{\bfsfG}({\bf k'},\omega)+ \omega {\rm Im}{\bfsfG}({\bf k'},\omega) \times {\bf k'} \times ({\dot {\bf R}} -\frac{\hbar{\bf k'}}{\gamma M})\nonumber \\
&&+ \omega {\dot {\bf R}} \times {\bf k'} \times {\rm Im}{\bfsfG}({\bf k'},\omega)\nonumber \\
&&+ {\dot {\bf R}} \times {\bf k'} \times {\rm Im}{\bfsfG}({\bf k'},\omega) \times {\bf k'} \times ({\dot {\bf R}} -\frac{\hbar{\bf k'}}{\gamma M}) \Big\} \cdot {\hat{\bf d}}_A \Big].
\end{eqnarray}
Here, the center-of-mass velocity of the atom before and after the emission of photon are, respectively, given by ${\dot {\bf R}}={\hbar{\bf q}}/{M}$ and ${\dot {\bf R}} - {\hbar{\bf k'}}/{\gamma M}$. Obviously, it is seen that the decay rate $\Gamma$ and the frequency shift $\delta \omega$ have been affected by the atomic motion
%
% the Eqs.~(\ref{relativistic decay rate}) and (\ref{relativistic Lamb shift}) are general and
through complicated relations that can be calculated by having the Green's tensor of any arbitrary medium, but due to the complexity of relations, its analytical calculations are practically impossible and should be investigated numerically. Now, in the limiting case where the surrounding environment of the moving atom is vacuum, the Green tensor of the system is written as follows
\begin{equation}\label{vacuum Green tensor}
{\rm Im}{\bfsfG}_{\alpha \beta}({\bf k},\omega) = (\delta _{\alpha \beta}-k_\alpha k_\beta) \frac{\pi}{2\omega c}\delta (|{\bf k} | - \frac{\omega}{c}).
\end{equation}
By substituting the above equation into Eq.~(\ref{relativistic decay rate}), as a well known result of special relativity, we find that $\Gamma = \frac{\omega _0^3 |{\bf d}_{0A}|^2}{3\pi \varepsilon _0\hbar c^3\gamma } = \frac{\Gamma _0}{\gamma }$, wherein ${\bf d}_{0A}$ and ${\Gamma _0}$ are, respectively, the transition dipole matrix element and the free space rate of decay of the atom in the rest frame of the atom.
This agreement also provides an independent check of the theoretical method introduced in this article.
\section{conclusions}\label{Sec:conclusions}
In this paper, a canonical quantization of the electromagnetic field interacting with moving charge particles is presented in the presence of an isotropic, homogeneous, and absorbing magnetodielectric medium. To achieve this purpose, an appropriate Lagrangian is introduced and the Hamiltonian of the combined system is derived. A unitary transformation has been applied on the Hamiltonian and the Hamiltonian in a multipolar form is obtained.
The formalism is generalized to include the relativistic motion of a atom, and then the evolution of the atomic system is determined in Schr\"{o}dinger picture.
By using the Markov approximation and finding the probability amplitudes of the system state vector, we got a general relation for the spontaneous emission rate and the Lamb shift of a moving atom with relativistic velocities in the presence of dissipation magnetodielectric mediums. It is shown that the dissipation effects of the surrounding environment on the spontaneous emission rate have been entered in our calculations through the Green tensor of the system, while the relativistic motion effects are introduced via the complicated relations including the parameter $\gamma$.
\section*{Reference}


\begin{thebibliography}{10}
%
\bibitem{Milonni1994}
P.W. Milonni, The quantum vacuum: an introduction to quantum electrodynamics, Academic Press, 1994. %Boston, Mass
%
\bibitem{Guo2008}
W. Guo, Phy. Rev. A. 77 (2008) 062111.
%
\bibitem{Wilkens1994}
M. Wilkens, Phys. Rev. A. 49 (1994) 570.
%
\bibitem{Boussiakou2002}
L.G. Boussiakou, C.R. Bennett, M. Babiker, Phys. Rev. Lett. 89 (2002) 123001.
%
\bibitem{Cresser2003}
J.D. Cresser, S.M. Barnett, J. Phys. B: At. Mol. Opt. Phys. 36 (2003) 1755.
%
\bibitem{Hinds2009}
E. Hinds, S.M. Barnett, Phys. Rev. Lett. 102 (2009) 050403.
%
\bibitem{Barnett2010a}
S.M. Barnett, R. Loudon, Proc. R. Soc. London, Ser. A. 368 (2010) 927.
%
\bibitem{Barnett2010b}
S.M. Barnett, Phys. Rev. Lett. 104 (2010) 070401.
%
\bibitem{Milonni2010} P.W. Milonni and R.W. Boyd, Adv. Opt. Photon. 2 (2010) 519.
%
\bibitem{Leonhardt1998}
U. Leonhardt, M. Wilkens, Europhys. Lett. 42 (1998) 365.
%
\bibitem{Leonhardt1999}
U. Leonhardt, P. Piwnicki, Phys. Rev. Lett. 82 (1999) 2426.
%
\bibitem{Horsley2005}
S. Horsley, M. Babiker, Phys. Rev. Lett. 95 (2005) 010405.
%
\bibitem{Ashkin2006}
A. Ashkin, Optical trapping and manipulation of neutral particles using lasers: a reprint volume with commentaries, World Scientific, 2006.
%
\bibitem{Scheel2009}
S. Scheel, S.Y. Buhmann, Phy. Rev. A. 80 (2009) 042902.
%
\bibitem{Barton2010}
G. Barton, New J. Phys. 12 (2010) 113045.
%
\bibitem{Lannebère2017}
S. Lannebère, M.G. Silveirinha, J. Opt. 19 (2017) 014004.
%
\bibitem{Amoghorban2010}
F. Kheirandish, E. Amoghorban, Phy. Rev. A. 82 (2010) 042901.
%
\bibitem{Amooghorban2011}
E. Amooghorban, M. Wubs, N.A. Mortensen, F. Kheirandish, Phys. Rev. A. 84 (2011) 013806.
%
\bibitem{Morshed2016}
M.M. Behbahani, E. Amooghorban, A. Mahdifar, Phys. Rev. A. 94 (2016) 013854.
%
\bibitem{Huttner1992}
B. Huttner, S.M. Barnett, Phys. Rev. A. 46 (1992) 4306.
%
\bibitem{Jeffers1996}
J. Jeffers, S.M. Barnett, R. Loudon, R. Matloob, M. Artoni, Opt Commun. 131 (1996) 66.
%
\bibitem{Suttorp2004}
L.G. Suttorp, M. Wubs, Phys. Rev. A. 70 (2004) 013816.
%
\bibitem{Amooshahi2009}
M. Amooshahi, J. Math. Phys 50 (2009) 062301.
%
\bibitem{Philbin2010}
T.G. Philbin, New J. Phys. 12 (2010) 123008.
%
\bibitem{Hopfield1996}
J.J. Hopfield, Phy. Rev. 112 (1996) 1555.
%
\bibitem{Dung2003}
H.T. Dung, S.Y. Buhmann, L. Kn\"{o}ll, D.-G. Welsch, J. K\"{a}stel, Phys. Rev. A. 68 (2003) 043816.
%
\bibitem{Buhmann2007}
S.Y. Buhmann, D.-G. Welsch, Prog. Quantum Electron. 31 (2007) 51.
%
\bibitem{Lembessis1993}
V.E. Lembessis, M. Babiker, C. Baxter, R. London, Phys. Rev. A. 48 (1993) 1594.
%
\bibitem{Guilot2002}
J.C. Guilot, J. Robert, J. Phys. A: Math. Gen. 35 (2002) 5023.
%
\bibitem{Bethe1957}
H.A. Bethe, E.E. Salpeter, Quantum mechanics of one- and two-electron atoms, Springer Science and Business Media, 1957. %Verlag Berlin
%
\bibitem{Buhmann2012a}
S.Y. Buhmann, Dispersion Forces I, Springer, 2013. %Verlag Berlin
%
\bibitem{Buhmann2012b}
S.Y. Buhmann, Dispersion Forces II, Springer, 2013. %Verlag Berlin
%
\bibitem{Amoghorban2014}
E. Amoghorban, F. Kheirandish, Internat. J. Theoret. Phys. 53 (2014) 2593.
%
\bibitem{Bogoliubov1980}
N.N. Bogoliubov, D.V.E. Shirkov, S. Chomet, Introduction to the Theory of Quantized Fields, John Wiley and Sons, New York, 1980. %Wiley-Interscience
%
\bibitem{Pendry1997} J. B. Pendry, J. Phys.: Condens. Matter 9 (1997) 10301.
%
\bibitem{Volokitin2006} A.I. Volokitin and B.N.J. Persson, Phys. Rev. B 74 (2006) 205413.
%
\bibitem{Philbin2009} T.G. Philbin and U. Leonhardt, New J. Phys. 11 (2009) 033035.
%
\bibitem{Pendry2010} J.B. Pendry, New J. Phys. 12 (2010) 033028.
%
\bibitem{Barton2010a} G. Barton, New J. Phys. 12 (2010) 113044.
%
\bibitem{Barton2010b} G. Barton, New J. Phys. 12 (2010) 113045.
%
\bibitem{Dedkov2002} G.V. Dedkov, and A.A. Kyasov, Phys. Solid State
44 (2002) 1809.
%
\bibitem{Golyk2013} V.A. Golyk, M. Kr¨uger and M. Kardar, Phys. Rev. B
88 (2013) 155117.
%
\bibitem{Pieplow2013} G. Pieplow and C. Henkel, New J. Phys. 15 (2013) 023027.
%
\bibitem{Intravaia2015}
F. Intravaia, V.E. Mkrtchian, S.Y. Buhmann, S. Scheel, D.A.R. Dalvit, C. Henkel, J. Phys.: Condens. Matter. 27 (2015) 214020.
%
\bibitem{Intravaia2016}
F. Intravaia, R.O. Behunin, C. Henkel, K. Busch, D.A.R. Dalvit, Phys. Rev. A. 94 (2016) 042114.
%
\bibitem{Klatt2017}
J. Klatt, M.B. Farías, D.A.R. Dalvit, S.Y. Buhmann, Phys. Rev. A. 95 (2017) 052510.
%
\end{thebibliography}
\end{document}